\documentclass[aps,prd,preprint,groupedaddress,nofootinbib,longbibliography]{revtex4-1}

\usepackage{amssymb}
\usepackage{amsmath}
\usepackage[linktocpage=true]{hyperref}
\usepackage{graphicx}

\begin{document}

\title{Free motion around black holes with discs or rings:\\
       between integrability and chaos -- VI. The Melnikov method}

\author{L. Polcar}
\email[]{polcar.vm@seznam.cz}
%\homepage[]{Your web page}
%\thanks{}
\affiliation{Institute of Theoretical Physics, Faculty of Mathematics and Physics,
             Charles University, Prague, Czechia}
\affiliation{Astronomical Institute,
             Czech Academy of Sciences,
             Ond\v{r}ejov, Czechia}

\author{O. Semer\'ak}
\email[]{oldrich.semerak@mff.cuni.cz}
%\homepage[]{Your web page}
%\thanks{}
\affiliation{Institute of Theoretical Physics, Faculty of Mathematics and Physics,
             Charles University, Prague, Czechia}

\date{\today}

\begin{abstract}
Motivated by black holes surrounded by accretion structures, we consider in this series static and axially symmetric black holes ``perturbed'' gravitationally as being encircled by a thin disc or a ring. In previous papers, we employed several different methods to detect, classify and evaluate chaos which can occur, due to the presence of the additional source, in time-like geodesic motion. Here we apply the Melnikov-integral method which is able to recognize how stable and unstable manifolds behave along the perturbed homoclinic orbit. Since the method standardly works for systems with one degree of freedom, we first suggest its modification applicable to two degrees of freedom (which is the our case), starting from a suitable canonical transformation of the corresponding Hamiltonian. The Melnikov function reveals that, after the perturbation, the asymptotic manifolds tend to split and intersect, consistently with the chaos found by other methods in the previous papers.
\end{abstract}

\maketitle

\section{Introduction}

In astrophysical models of accreting black holes, the gravitational effect of the accreting material is usually neglected, hence the space-time is being described by the Kerr or Schwarzschild metric, corresponding to {\em isolated} stationary black holes. Such an approximation is certainly justified on the level of potential (i.e. metric), but in the vicinity of the outer matter it may well fail for the field and mainly for higher derivatives of the potential (space-time curvature). One of clear consequences of the perturbation is that the geodesic dynamics looses complete integrability (originally valid in the field of {\em isolated} stationary and axisymmetric black holes, \citealt{FrolovKK-17}). Hence, one of suitable ways how to examine the deviation of space-time from the Kerr or Schwarzschild ideal (and then possibly its observational implications) is to study the geodesic motion by methods of dynamical-system theory \citep{GairLM-08,LukesGAC-10}.

In this series of papers, we have employed various methods to detect, classify and evaluate chaos occurring in the geodesic motion in the above backgrounds, focusing on those generated by a static black hole encircled by an axially symmetric thin disc or ring. We will not repeat all the results obtained, let us only refer to the last two papers, \cite{WitzanySS-15} where we compared the results with those obtained by a Newtonian treatment of the corresponding pseudo-Newtonian system, and \cite{PolcarSS-19} where we compared the effect of the Bach-Weyl ring on the Schwarzschild black hole with that of the Majumdar-Papapetrou ring on the extreme Reissner-Nordstr\"om black hole, and also tested one of the curvature-based criteria for chaos.

The present paper slightly differs from the previous ones, because it concerns an {\em analytical} method -- that of the Melnikov integral -- and  because it mainly focuses on the method itself, namely, we suggest its modification suitable for systems with two degrees of freedom (in a standard version, the method applies to only one degree of freedom). By applying the method to our specific system, we then further support the results obtained by previous methods and confirm that the geodesic chaos observed around perturbed black holes {\em is} of homoclinic origin.

We start (section \ref{Melnikov-classical}) by a minimal summary of the classical Melnikov method. In section \ref{Hamiltonians}, Hamiltonians are written down describing our pseudo-Newtonian and relativistic geodesic systems. In section \ref{Hamiltonians}, we perform a canonical transformation to a suitable action-angle coordinates in order to adapt the Melnikov method to our systems with two degrees of freedom. Hyperbolic fixed points (unstable circular orbits) and the attached homoclinic orbits of the systems are found then in section \ref{homo-orbits} and the corresponding Melnikov-function integral is drawn up in section \ref{Melnikov}. The latter is evaluated numerically in section \ref{numerics} and checked against the numerically integrated geodesic flow (portrayed on Poincar\'e diagrams), including a few comments on the Melnikov method as such. Finally, we summarize our results and mention some literature in section \ref{concluding}.

Not to repeat the introduction on static and axially symmetric space-times and specifically on sources we consider here (Schwarzschild black hole, extreme Reissner-Nordstr\"om black hole, Bach-Weyl ring, inverted first Morgan-Morgan counter-rotating disc, Majumdar-Papapetrou ring), we ask the reader to see the previous papers of this series, of which we explicitly refer to \cite{WitzanySS-15} and \cite{PolcarSS-19}.

We basically use standard notation, with $g_{\mu\nu}$ being the metric tensor (of $-$$+$$+$$+$ signature), $M$ denoting mass of the central black hole, ${\cal M}$ denoting mass of the exterior source (ring or disc) and $b$ its radius. Following usual notation, we call the Melnikov function $M(\vartheta_0)$. This will not cause confusion (with the black-hole mass $M$), because the Melnikov function will always be written together with the variable $\vartheta_0$ indicating the homoclinic orbit in terms of a suitably defined angle (and it will anyway be clear from the context).

\subsection{Weyl solutions and their Majumdar-Papapetrou subclass}

Since we speak of motivation by accreting black holes, it is at place to add a note concerning the sources we employ. The background space-times we considered in all previous papers of this series, as well as in the present one, belong to the Weyl class which, in the Weyl cylindrical-type coordinates $(t,\rho,z,\phi)$, can be described by the metric
\begin{equation}  \label{Weyl-metric}
  {\rm d}s^2=-e^{2\nu}{\rm d}t^2+\rho^2 e^{-2\nu}{\rm d}\phi^2
             +e^{2\lambda-2\nu}({\rm d}\rho^2+{\rm d}z^2) \,,
\end{equation}
where the unspecified functions $\nu$ and $\lambda$ only depend on cylindrical-type radius $x^1\!\equiv\!\rho$ and the ``vertical" linear coordinate $x^2\!\equiv\!z$. These two coordinates cover, in an isotropic manner, the meridional planes which are everywhere orthogonal to the Killing planes, spanned by the two existing Killing symmetries (stationarity and axisymmetry) and covered by the respective adapted coordinates $t$ (time) and $\phi$ (azimuth). The above metric holds provided that the energy-momentum tensor satisfies $T^1_1+T^2_2=0$. The Einstein equations imply that the potential $\nu$ satisfies the Laplace/Poisson equation, while $\lambda$ can be obtained from (an already known) $\nu$ by line integration.

In the first papers of this series, we considered, as a limiting approximation of an accretion structure, the thin ring described by the Bach-Weyl solution; this is a direct counter-part of the ordinary homogeneous Newtonian ring. Despite its potential $\nu$ is taken over from the Newtonian treatment, in general relativity such a ring generates a surprisingly weird geometry in its vicinity due to the second metric function $\lambda$ \citep{Semerak-16}. In order to check whether the chaos induced in such a system is not just induced by this undesired feature, we rather considered, in the last paper \citep{PolcarSS-19}, a different ring whose field does not suffer such a pathology -- the extremally charged, Majumdar-Papapetrou ring (around an extreme Reissner-Nordstr\"om black hole) -- and compared the results with those obtained for the Bach-Weyl ring. We already explained this point in the preceding paper, but let us repeat that this is by no means to say that black-hole accretion systems are extremally charged. However, for the study of {\em geodesic} motion (i.e. that of free {\em uncharged} particles), the chargedness of the sources is actually an advantage, since their electrostatic field (in the Majumdar-Papapetrou case having the same shape as the gravitational one) can mimic the diluted matter, likely present in real systems.

More specifically, the Majumdar-Papapetrou solutions form a subclass of the Weyl solutions determined by $\lambda\!=\!0$ and by the relation $\Phi\!=\!e^\nu$ between their electrostatic potential $\Phi$ and the lapse function $N\!\equiv\!e^\nu$. In other words, their four-potential reads $A_\mu=(-e^\nu,0,0,0)$ in the Weyl coordinates, which means that the electromagnetic-field tensor has but two non-trivial components $F_{ti}=e^\nu\nu_{,i}\,$, so the electrostatic invariant is
\[F_{\mu\nu}F^{\mu\nu}=-2e^{2\nu}\left[(\nu_{,\rho})^2+(\nu_{,z})^2\right]\]
and the non-zero components of the Ricci tensor $R_{\mu\nu}$ ($=8\pi T_{\mu\nu}$ from Einstein's equations) read
\begin{equation}
  -R^t_t=R^\phi_\phi = e^{2\nu}\left[(\nu_{,\rho})^2+(\nu_{,z})^2\right],
  \qquad
  -R_{\rho\rho}=R_{zz} = (\nu_{,\rho})^2-(\nu_{,z})^2,
  \qquad
  R_{\rho z} = -2\nu_{,\rho}\nu_{,z} \;.
\end{equation}
Any observer at rest, with four-velocity $u^\mu=(e^{-\nu},0,0,0)$, measures the electric field $E_i=F_{i\beta}u^\beta=-\nu_{,i}$ (and no magnetic one) and the electromagnetic energy density
\[T_{\alpha\beta}u^\alpha u^\beta=\frac{R_{tt}}{8\pi}\,e^{-2\nu}=\frac{-R^t_t}{8\pi}
                                 =\frac{e^{2\nu}}{8\pi}\left[(\nu_{,\rho})^2+(\nu_{,z})^2\right]
                                 =\frac{1}{8\pi}\,g^{ij}E_i E_j \;.\]

\section{Classical Melnikov method}
\label{Melnikov-classical}

The Melnikov method detects how a flow of a dynamical system behaves, under perturbation, in the vicinity of its homoclinic orbit. Homoclinic orbit is a (closed) trajectory which has a hyperbolic (saddle) fixed point of the respective flow as both its past and future asymptote. Along such an orbit, the tangent bundle of the configuration manifold splits into stable and unstable invariant subbundles, spanning submanifolds in which the flow is contracting and expanding, respectively (thus {\em saddle} point). In the unperturbed system, these submanifolds intersect ``longitudinally'' (thus coincide) along the homoclinic orbit. If the perturbation deforms them in such a way that they start to intersect transversally along that orbit, it is a signature that the orbit has ``broken up'' into a chaotic layer (so-called Smale-Birkhoff homoclinic theorem). The necessary theory is explained in \cite{Wiggins-03}, for instance.

It is a salient feature of the geodesic flow in the space-times of stationary black holes that it does contain hyperbolic fixed points: the latter are represented by unstable periodic (spatially circular) geodesics. The attached homoclinic orbits are those which asymptotically ``unwind'' from them, make a ballistic loop and ``wind back'' in a symmetric manner.\footnote
{The homoclinic orbits are also often called separatrices, because they typically separate two distinct types of evolution. In the black-hole case, in particular, they separate the ``eternal'' bound orbits from those which plunge into the horizon.}
To imagine, intuitively, the stability properties at the above circular orbits, one realizes that the time and azimuthal directions are neutral since they are being ``held'' by the respective two constants of geodesic motion -- energy and angular momentum with respect to infinity; the latitudinal direction is stable since the particle shifted off the equatorial plane is being pulled back; and, finally, the radial direction is unstable -- that is the direction in which the geodesic flow primarily diverges. Therefore, in the stationary black-hole case, it is the $(r,u^r)$ plane of the geodesic phase space which is interesting and which is thus being mainly studied (used for plotting the Poincar\'e diagrams, etc.).

We now return to a general dynamical system and briefly sketch the standard version of the Melnikov method.
Consider a one-degree-of-freedom system whose Hamiltonian is at least a $C^2$ function and can be expressed as
\begin{equation}  \label{H0,H1}
  H(q,p,t;\epsilon)=H_0(q,p)+\epsilon H_1(q,p,t)+{\cal O}(\epsilon^2),
\end{equation}
where $[q,p]$ is a phase-space point, $t$ is time and $\epsilon\!>\!0$ is a small parameter.
In addition, let us assume it has the following properties:\\
(i) $H_1$ is a periodic function of $t$, with some period $T$;\\
(ii) $H_0$ has a hyperbolic fixed point $[Q_0,P_0]$, connected to itself by a homoclinic orbit $[q_0(t),p_0(t)]$, $\lim\limits_{t\rightarrow\pm\infty} [q_0(t),p_0(t)]=[Q_0,P_0]$.\\
The picture is slightly changed after ``switching on'' the perturbation $H_1$, with a crucial question being whether the asymptotic manifolds, originally coinciding along the homoclinic orbit, now intersect transversally there. Exactly this point is addressed by the Melnikov method, namely, the method computes a distance between the (perturbed) stable and unstable manifolds at the unperturbed homoclinic orbit. For a one-degree-of-freedom system (system confined to one {\em position} dimension), the phase space is three-dimensional and the unperturbed homoclinic manifold (represented by the coinciding asymptotic manifolds) is two-dimensional, so one can take its normal and compute the distance between intersections of that normal with the {\em perturbed} asymptotic manifolds.

It has been shown (see e.g. \citealt{Wiggins-03} again) that in the first order of perturbation ($\epsilon$) the above distance is proportional to a function which can be expressed as an integral, along the unperturbed homoclinic orbit, of the Poisson bracket of $H_0$ with $H_1$,
\begin{equation}
  M(t_0)=\int\limits_{-\infty}^{+\infty}\left\{H_0,H_1\right\}(q_0(t),p_0(t),t_0+t)\,{\rm d}t \,,
\end{equation}
where $t_0$ is some chosen value of time. This function, called Melnikov's function, is periodic, has the same period $T$ as $H_1$, and the transverse intersection of the perturbed asymptotic manifolds happens when
\begin{equation}
  M(t_0)\!=\!0
  \qquad {\rm and} \qquad
  \frac{{\rm d}M}{{\rm d}t_0}(t_0)\neq 0 \,.
\end{equation}
(It also holds that if $M$ has no zeros, then the asymptotic manifolds do not intersect at all.)
Note that the second condition is necessary, because it excludes the case when $M$ is zero {\em identically} which would mean that the asymptotic manifolds keep coinciding (at the homoclinic orbit) even after the perturbation.

To summarize the technique, one has to know the Hamiltonian of a dynamical system, split it according to (\ref{H0,H1}), find the homoclinic orbit (if there exists some) by solving the unperturbed equations of motion, compute the Poisson bracket between $H_0$ and $H_1$ and integrate it along the unperturbed homoclinic orbit. If the result satisfies the above two conditions (has ``simple zeros''), the perturbation makes the dynamics chaotic close to the original separatrix. If there are no simple zeros, the homoclinic chaos does not occur. 

Let us stress again that the above version of the Melnikov method is restricted to systems with one degree of freedom, whereas our system has two degrees of freedom. Actually, in space-times with two commuting Killing symmetries, stationarity and axial symmetry, one always has two constants of geodesic motion, energy and axial angular momentum, which determine (respectively) the time and azimuthal components of four-momentum (in coordinates adapted to those symmetries). The remaining freedom is bound to the meridional planes, i.e. the surfaces everywhere orthogonal to both symmetries.\footnote
{It is {\em not} automatic that such planes exist as integral (global) submanifolds of space-time, so actually one has to {\em assume} that. Such a property is called orthogonal transitivity and is equivalent to the situation when the sources only follow spatially circular orbits (such space-times are thus called {\em circular} space-times) with steady angular velocity, i.e. their motion points in a direction given by combination of the Killing vectors.}
Usually these are covered by either cylindrical-type coordinates (e.g. of the Weyl type, $\rho$ and $z$) or spheroidal coordinates (e.g. of the Schwarzschild type, $r$ and $\theta$). In order to adapt the Melnikov method to our problem, we will reformulate it using the approach suggested by \cite{HolmesM-83}. However, we will have to make a canonical transformation in order to be able to put the Hamiltonians into the appropriate form. Below, we first derive the original Hamiltonians for a pseudo-Newtonian as well as relativistic formulation of our problem.

\section{Hamiltonians for pseudo-Newtonian and relativistic systems}
\label{Hamiltonians}

In this series of papers, we have been studying geodesic dynamics in two types of black-hole space-times, deformed by the presence of a gravitating ring or disc. Since we restrict to static and axisymmetric (electro-)vacuum situation, the first metric function (potential) $\nu$ always superposes linearly. For the second function $\lambda$ this is not in general the case. Actually, for the Schwarzschild black hole encircled by some additional source,\footnote
{We have specifically been considering the Bach-Weyl ring, the inverted first Morgan-Morgan disc and one of the discs with power-law density profile in this series.}
$\lambda$ has to be found numerically. Therefore, it is not easy to study these configurations {\em analytically}, at least concerning those properties in which $\lambda$ is relevant. Since the geodesic dynamics {\em is} such a property ($\lambda$ is needed for Christoffel symbols), we will resort to its pseudo-Newtonian treatment in those cases. On the contrary, for the extreme Reissner-Nordstr\"om black hole encircled by the Majumdar-Papapetrou ring (suggested for study in \citealt{PolcarSS-19}), $\lambda\!=\!0$, so the analytical approach is relatively easy. Let us write down Hamiltonians for both cases.

In the Newtonian treatment, the Hamiltonian for a test particle of mass $m$ in the potential $V_{\rm BH}$ of a ``black hole'' plus that due to the another source ($\nu_{\rm ext}$) reads
\begin{equation}
  H(r,\theta,p_i)=\frac{p^2}{2m}+mV_{\rm BH}(r)+m\nu_{\rm ext}(r,\theta) \,.
\end{equation}
It can be decomposed as
\begin{equation}  \label{H,Newt}
  H(r,\theta,p_i)=H_0(r,\theta,p_i)+\frac{{\cal M}}{M}\,H_1(r,\theta)
\end{equation}
into an integrable part 
\begin{equation}  \label{H0,Newt}
  H_0(r,\theta,p_i)
  =\frac{1}{2m}\left[p_r^2+\frac{1}{r^2}\left(p_\theta^2+\frac{p_\phi^2}{\sin^2\theta}\right)\right]
   +m V_{\rm BH}(r)
\end{equation}
and the perturbation
\begin{equation}  \label{H1,Newt}
  H_1(r,\theta)=\frac{mM}{{\cal M}}\,\nu_{\rm ext}(r,\theta) \,,
\end{equation}
where the small perturbation parameter ($\epsilon$) is in our case represented by the relative mass ${\cal M}/M$ of the exterior source with respect to the black-hole mass $M$. Note that the exterior potential $\nu_{\rm ext}$ is proportional to ${\cal M}$, so $H_1$ actually does not depend on ${\cal M}$.

\begin{figure}[h]\centering
\includegraphics[width=130mm]{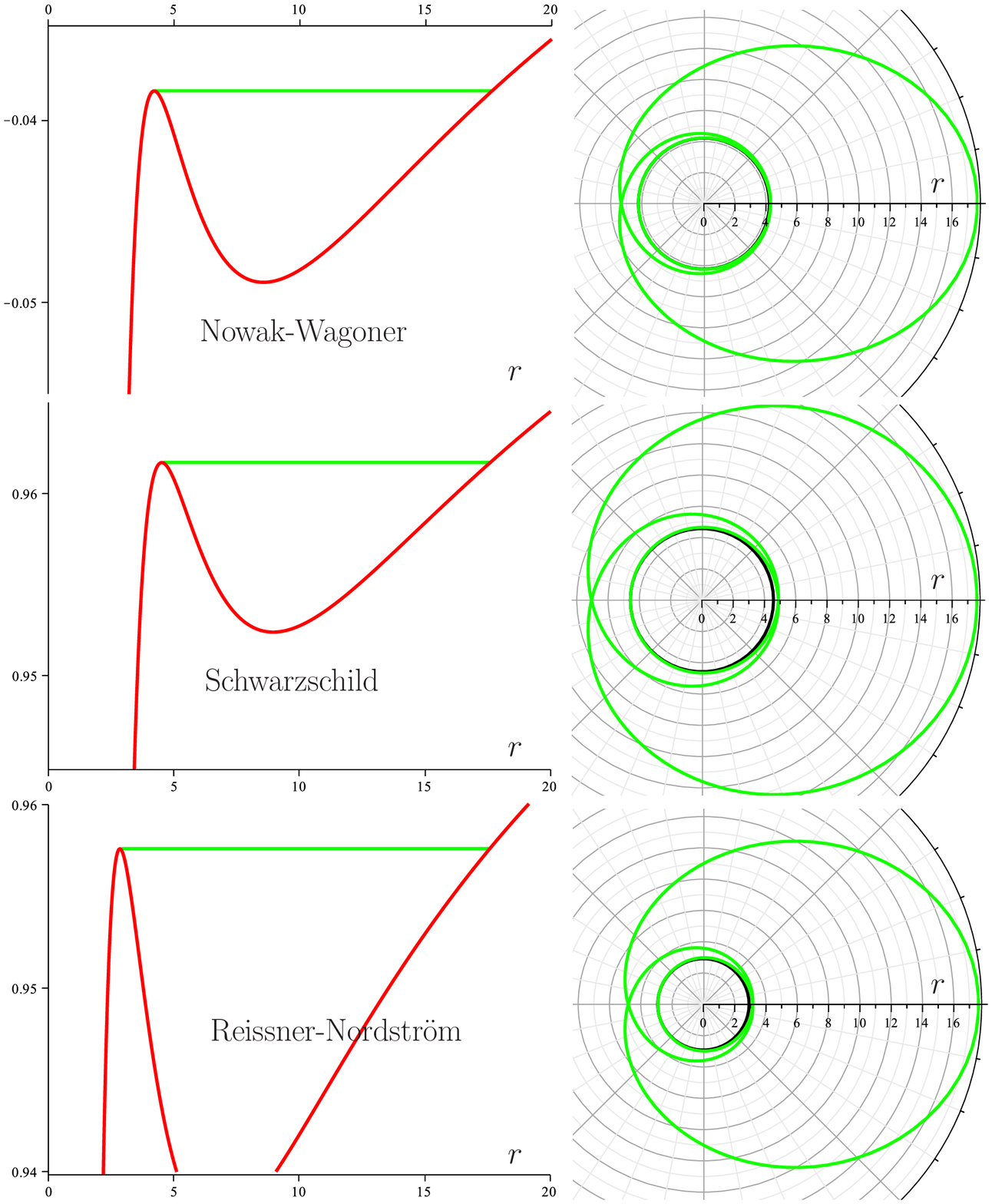}
\caption{Examples of homoclinic orbits of the Nowak-Wagoner pseudo-potential (top row), of the Schwarzschild space-time (middle row), and of the Reissner-Nordstr\"om space-time (bottom row). In the left column, radial shapes of the effective potentials are drawn in red with the homoclinic-orbit energy levels indicated in green, while in the right column, spatial shapes of these orbits are shown in green (the unstable circular orbits which they have as both past and future asymptotes are also plotted in black, but are barely visible, even though we only draw $-3\pi\leq\vartheta\leq 3\pi$ part of the homoclinic orbits). For easy comparison, we select orbits with the same maximal radius (apocentre) of $r\!=\!17.6M$; this corresponds to $\ell\!=\!2.6M$ for Nowak-Wagoner, to $\ell\!=\!3.669M$ for Schwarzschild and to $\ell\!=\!3.091M$ for Reissner-Nordstr\"om.}
\label{homoclinic-orbits}
\end{figure}

What remains to be decided is how to mimic the actual black-hole field. Several different ``pseudo-Newtonian'' potentials have been suggested for this purpose in the literature. In one of previous papers, by \cite{WitzanySS-15}, we provided, together with relevant references, a review of some of them (including a one we newly suggested), and tested numerically how well the corresponding Keplerian-motion dynamics resembles the exact relativistic one. Although this comparison did not come out very well for the Nowak-Wagoner potential
\begin{equation}  \label{NW-potential}
  V_{\rm BH}\equiv V_{\rm NW}=-\frac{M}{r}\left(1-\frac{3M}{r}+\frac{12M^2}{r^2}\right),
\end{equation}
we will adhere to the latter in the present paper, because for this potential it is quite easy to find the homoclinic orbit explicitly (and the result is quite similar to its exact relativistic counterpart, see Fig. \ref{homoclinic-orbits}).

One more remark to the pseudo-Newtonian approach.
There necessarily arises the following question:
which coordinates covering the curved relativistic space-time are adequate counter-parts of Euclidean coordinates of the (pseudo-)Newtonian description? To answer this question, it is important to add that our exterior source will be a ring or a disc, natively described by the Weyl-type metric in Weyl-type cylindrical coordinates.
Since the issue of coordinates was discussed in \cite{WitzanySS-15}, we just follow our recommendation from there:
(i) take the black-hole pseudo-potential, as originally expressed in the Euclidean spherical coordinates $r$, $\theta$, and consider that it should imitate the black hole described in the Schwarzschild coordinates; (ii) take the potential of the disc or a ring, originally expressed in Euclidean cylindrical coordinates ($\rho$, $z$), and realize that it corresponds, in the relativistic description, to a disc or a ring potential represented in Weyl coordinates (because in them it is determined by the same equation, namely the Laplace one); hence, (iii) add these two potentials after transforming the disc/ring potential to the spheroidal coordinates ($r$, $\theta$) according to the relations valid between the Weyl and the Schwarzschild-type coordinates, i.e.
\begin{equation}  \label{rho,z-Weyl}
  \rho=\sqrt{r(r-2M)}\,\sin\theta, \qquad z=(r-M)\cos\theta
\end{equation}
or
\begin{equation}  \label{rho,z-Weyl,extreme}
  \rho=(r-M)\sin\theta, \qquad z=(r-M)\cos\theta,
\end{equation}
where the second form applies to the case involving {\it extreme} black hole. 
(We will use the first transformation when making superpositions with the Schwarzschild black hole, while the second transformation for superpositions with the extreme Reissner-Nordstr\"om black hole.)

Now to the relativistic version of the geodesic problem. It is described by the Hamiltonian
\begin{equation}
  H(r,\theta,p_\alpha)=\frac{1}{2m}\,g^{\mu\nu}(r,\theta)\,p_\mu p_\nu \,,
\end{equation}
which can be expanded, in the small parameter ${\cal M}/M$, as
\begin{align}
  H(r,\theta,p_\alpha)
    &= \frac{1}{2m}\,p_\mu p_\nu
       \left[g^{\mu\nu}({\cal M}\!=\!0)
             +{\cal M}\,\frac{\partial g^{\mu\nu}}{\partial {\cal M}}({\cal M}\!=\!0)
             +{\cal O}({\cal M}^2)\right]  \nonumber \\
    &= H_0(r,\theta,p_\alpha)+\frac{\cal M}{M}\,H_1(r,\theta,p_\alpha)
       + {\cal O}({\cal M}^2),  \label{H,rel}
\end{align}
where, assuming that the metric is diagonal, one can take (for every fixed $\mu$ and $\nu$)
\begin{equation}  \label{metric,expansion}
  g^{\mu\nu}({\cal M}\!=\!0)=\frac{1}{g_{\mu\nu}({\cal M}\!=\!0)} \,,
  \qquad
  \frac{\partial g^{\mu\nu}}{\partial {\cal M}}({\cal M}\!=\!0)
  =\frac{-\frac{\partial g_{\mu\nu}}{\partial {\cal M}}}{(g_{\mu\nu})^2}({\cal M}\!=\!0) \,.
\end{equation}
For the Schwarzschild black hole, one has explicitly
\begin{equation}  \label{H0,Schw}
  H_0(r,\theta,p_\alpha)
    =\frac{1}{2m}\left[-\frac{p_t^2}{1-\frac{2M}{r}}+\left(1-\frac{2M}{r}\right)p_r^2
                       +\frac{1}{r^2}\left(p_\theta^2+\frac{p_\phi^2}{\sin^2\theta}\right)\right],
\end{equation}
while for the extreme Reissner-Nordstr\"om black hole
\begin{equation}  \label{H0,eRN}
  H_0(r,\theta,p_\alpha)
    = \frac{1}{2m}\left[-\frac{p_t^2}{\left(1-\frac{M}{r}\right)^{\!2}}+\left(1-\frac{M}{r}\right)^{\!2} p_r^2
                        +\frac{1}{r^2}\left(p_\theta^2+\frac{p_\phi^2}{\sin^2\theta}\right)\right].
\end{equation}

Comparing the Hamiltonians (\ref{H,Newt}) and (\ref{H,rel}) with the form (\ref{H0,H1}) necessary for the Melnikov method, it is clear that there are at least two problems: we have more degrees of freedom and our perturbations are {\em not} time dependent (of course: we restrict to {\em static} and axially symmetric configurations). In the following section, we fix this problem using the method suggested by \cite{HolmesM-83} and performing a suitable canonical transformation to action-angle coordinates.

\section{Modification of the Melnikov method}
\label{modification}

\cite{HolmesM-83} considered, in section 6 of their paper, the Hamiltonian
\begin{equation}  \label{H0,H1,2DF}
  H(q,p,\psi,J)=H_0(q,p,J)+\epsilon H_1(q,p,\psi,J)+{\cal O}(\epsilon^2)
\end{equation}
which describes a system with two degrees of freedom, does not depend explicitly on time, but one of its coordinate variables ($\psi$) is periodic. The idea is to use {\em this} variable in the role of time, with its conjugate momentum $J$ playing then the role of the Hamiltonian. Using the relation between $\psi$ and time $t$ obtained from Hamilton equations, i.e., for the unperturbed system,
\begin{equation}
  \dot\psi:=\frac{{\rm d}\psi}{{\rm d}t}=\frac{\partial H_0}{\partial J}(q,p,J),
\end{equation}
one can reparametrize the flow -- and the homoclinic orbit, in particular -- by $\psi$.

Evolution of the perturbed system is confined to some energy hypersurface $H(q,p,\psi,J)\!=\!{\rm const}\!=:\!h$, and if this equation is invertible, one can express from it $J\!=\!J(q,p,\psi,h)$. As Holmes \& Marsden showed, in terms of this new ``Hamiltonian'' the Melnikov function can be rewritten as
\begin{equation}  \label{Melnikov,new}
  M(\psi_0)=\int\limits_{-\infty}^{\infty}
            \frac{1}{\dot\psi\left(q_0(\psi),p_0(\psi),J\right)}
            \left\{H_0,\frac{H_1}{\dot\psi}\right\}\left(q_0(\psi),p_0(\psi),\psi+\psi_0,J\right)\,{\rm d}\psi \,,
\end{equation}
where $q_0(\psi),p_0(\psi)$ represent the homoclinic orbit for some fixed value of $J$ (the Poisson bracket is thus computed in the variables $q,p$ only). If, for a given value of $J$, the system has only one homoclinic orbit, then fixing $J$ is equivalent to fixing the energy surface $h\!=\!H_0(q_0(\psi),p_0(\psi),J)$.

The theorem about the occurrence of transverse intersections of asymptotic manifolds now reads as follows \citep{HolmesM-83}:
Let $H_0(q,p,J)$ have, for some fixed value of $J$, a hyperbolic fixed point and a homoclinic orbit $q_0(\psi),p_0(\psi)$ attached to it, and let $\dot\psi(q_0(\psi),p_0(\psi),J)>\!0$. Then, if $M(\psi_0)$ has simple zeros, the Hamiltonian (\ref{H0,H1,2DF}) (for $\epsilon>0$ sufficiently small) describes a system whose asymptotic manifolds intersect transversally on the energy surface $H\!=\!H_0(q_0(\psi),p_0(\psi),J)$. Note that ``energy'' ($h$) in fact means any integral of motion of the complete Hamiltonian; in our case, it will be the energy $E\!=\!-p_t$ and the azimuthal angular momentum $L_\phi\!=\!p_\phi$.\footnote
{The ``axial'' angular momentum (due to azimuthal motion) is mostly being denoted by $L$ and we also did it so in preceding papers of this series. Here, however, we will also need to introduce the second component of the angular momentum, so it is natural to use the present notation.}
Fixing these integrals of motion and the homoclinic orbit (if it does exist for some $E$ and $L_\phi$) automatically fixes the value of $J$.

Unfortunately, it is not directly possible to apply the above procedure to our perturbed black-hole case. Namely, in that case, symmetrical in $t$ and $\phi$, the variables are identified as $q\!\equiv\!r$, $p\!\equiv\!p_r$, $\psi\!\equiv\!\theta$, $J\!\equiv\!p_\theta$, where, however, $p_\theta$ is {\em not} an integral of motion for the unperturbed system ($H_0$ depends on $\theta$), so its conjugate coordinate $\theta$ cannot play the role of the above $\psi$. Nevertheless, let us show that it is possible to perform a canonical transformation in the second pair of variables, $(\theta,p_\theta)\rightarrow(\vartheta,J_\vartheta)$, such that the new momentum $J_\vartheta$ will be an integral of the unperturbed motion (it will be action-type variable), and its conjugate angle coordinate $\vartheta$ {\em can} play the role of $\psi$ from the Holmes-Marsden method.

Looking for a suitable canonical transformation, we first notice that all our unperturbed Hamiltonians, (\ref{H0,Newt}), (\ref{H0,Schw}) and (\ref{H0,eRN}), contain the same angular part, given by the variables $\theta$, $p_\theta$ and representing square of the total angular momentum,
\begin{equation}  \label{L^2}
   L^2:=p_\theta^2+\frac{p_\phi^2}{\sin^2\theta}\equiv p_\theta^2+\frac{L_\phi^2}{\sin^2\theta} \;.
\end{equation}
Since $L$ {\em is} an integral of motion for our unperturbed systems (because they are spherically symmetric), it is possible to compute the action variable $J_\vartheta$ using the standard definition by integral
\begin{equation}
  J_\vartheta=\frac{1}{2\pi}\oint p_\theta\,{\rm d}\theta
             =\frac{1}{\pi}\int\limits_{\theta_{\rm min}}^{\theta_{\rm max}}
              \sqrt{L^2-\frac{L_\phi^2}{\sin^2\theta}}\;{\rm d}\theta
\end{equation}
taken over the whole period between turning points given by $p_\theta\!=\!0$ and reading, from (\ref{L^2}),
\[\theta_{\rm min}=\arcsin\frac{L_\phi}{L} \,,
  \qquad
  \theta_{\rm max}=\pi-\theta_{\rm min} \,.\]

Now, since the Hamilton-Jacobi equation is separable for the unperturbed system,\footnote
{This even holds for motion of charged test particles in more general, stationary black-hole space-times -- see \cite{FrolovKK-17} for a recent review.}
i.e. the unperturbed action $S$ can be written as a sum $S=\sum_{i=1}^{N}S_i(q_i,p_1,...,p_N)$ ($N$ is the number of degrees of freedom), one has $p_\theta\!=\frac{\partial S_\theta}{\partial \theta}$ and thus
\begin{equation}
  J_\vartheta=\frac{1}{\pi}\int\limits_{\theta_{\rm min}}^{\theta_{\rm max}}
              \sqrt{L^2-\frac{L_\phi^2}{\sin^2\theta}}\;{\rm d}\theta
             =\frac{1}{\pi}\int\limits_{\theta_{\rm min}}^{\theta_{\rm max}}
              \frac{\partial S_\theta}{\partial\theta}(\theta,L,L_\phi)\,{\rm d}\theta \,.
\end{equation}
The primitive function of $\sqrt{L^2-\frac{L_\phi^2}{\sin^2\theta}}$ reads, up to a constant,
\begin{equation}
  S_\theta=\frac{L_\phi}{2}\;\arctan\frac{L^2\cos\theta+L^2-L_\phi^2}
                                         {L_\phi\,\sqrt{L^2\sin^2\theta-L_\phi^2}}
           +\frac{L_\phi}{2}\;\arctan\frac{L^2\cos\theta-L^2+L_\phi^2}
                                          {L_\phi\,\sqrt{L^2\sin^2\theta-L_\phi^2}}
           -L\,\arctan\frac{L\cos\theta}{\sqrt{L^2\sin^2\theta-L_\phi^2}} \;,      
\end{equation}
so, taking the limits to the turning points, one obtains a simple result
\begin{equation}
  J_\vartheta
     =\frac{1}{\pi}
      \left[S_\theta(\theta\!\to\!\theta_{\rm max})-S_\theta(\theta\!\to\!\theta_{\rm min})\right]
     =\frac{1}{\pi}\left(-\frac{\pi}{4}\,L_\phi-\frac{\pi}{4}\,L_\phi+\frac{\pi}{2}\,L\right)
      -\frac{1}{\pi}\left(\frac{\pi}{4}\,L_\phi+\frac{\pi}{4}\,L_\phi-\frac{\pi}{2}\,L\right)
     =L-L_\phi \,.
\end{equation}
This is indeed an integral of the motion. In order to find its conjugate coordinate $\vartheta$, we can use the transformation equation valid with the appropriate type of the canonical-transformation generating function (namely the one depending on old coordinates and new momenta),
\begin{equation}
  \vartheta=\frac{\partial S_\theta}{\partial J_\vartheta}(\theta,L_\phi,J_\vartheta)
           =\frac{\partial S_\theta}{\partial L}(\theta,L_\phi,L).
\end{equation}
(We are only transforming the coordinates $\theta$ and $p_\theta$, so it is really sufficient to include just $S_\theta$ part of the action.) Inverting the above equation, we can express $\theta$ as a function of $\vartheta$, which yields the final transformation relations
\begin{equation}  \label{canonical-trans}
  \theta=\pi-\arccos\left(\sqrt{1-\frac{L_\phi^2}{L^2}}\;\sin\vartheta\right),
  \qquad
  L=J_\vartheta+L_\phi \;.
\end{equation}

Finally, we express our Hamiltonians in terms of these new variables. For example, the unperturbed pseudo-Newtonian Hamiltonian now takes the form
\begin{equation}
  H_0(r,p_r,J_\vartheta,L_\phi)=\frac{1}{2m}\left[p_r^2+\frac{(J_\vartheta+L_\phi)^2}{r^2}\right]+mV_{\rm NW}(r) \,.
\end{equation}
The relativistic Hamiltonians are similar since they also contain the angular-momentum part $\frac{(J_\vartheta+L_\phi)^2}{r^2}\,$.
This is already the desired form, namely that of (\ref{H0,H1,2DF}) with $\vartheta$ playing the role of $\psi$, since (i) $H_0$ now does not depend on the angular coordinate ($\vartheta$), while the angular momenta $J_\vartheta$ and $L_\phi$ are treated on equal footing; and (ii) in $H_1$, the variable $\theta$ has been replaced by $\vartheta$. The complete Hamiltonian then reads
\begin{equation}
  H(r,p_r,\vartheta,J_\vartheta)=H_0(r,p_r,J_\vartheta)+mH_1(r,p_r,\vartheta,J_\vartheta)+{\cal O}(\epsilon^2) \,,
\end{equation}
where we have not listed, as variables, the additional momenta $p_t\!\equiv\!-E$ and $p_\phi\!\equiv\!L_\phi$ since they are just parameters.

Before concluding this section, note that the term $\frac{\partial S}{\partial t}$ in the Hamiltonian's canonical transformation
\[\bar{H}(\bar{q}_i,\bar{p}_i,t)=H(q_i,p_i,t)+\frac{\partial S(q_i,\bar{p}_i,t)}{\partial t}\]
only adds a constant and so we have omitted it in the above Hamiltonians. Also note that, although $J_\vartheta$ is an action variable, the pair $(\vartheta,J_\vartheta)$ does {\em not} represent action-angle coordinates, because $H_0$ still depends on $r$. Nevertheless, for our purposes it is sufficient that the Hamilton equations in $(\vartheta,J_\vartheta)$ are equivalent to those expressed in $(\theta,p_\theta)$.

%%%%%%%%%%%%%%%%%%%%%%%%%%%%%%%%%%%%%%
%%%%%%%%%%%%%%%%%%%%%%%%%%%%%%%%%%%%%%

\section{Homoclinic orbits}
\label{homo-orbits}

In order to compute the Melnikov integral, we first need to find the integration path -- the homoclinic orbit (separatrix) of the unperturbed system. For the black-hole space-times we are interested in, such orbits have been studied in several papers, of which we refer to \cite{LevinPG-08} for a thorough and clear description of what they are and what is their place within the geodesic flow. The nature of the homoclinic orbit is best seen from the effective-potential graph (see below): it has the same energy (and angular momentum) as the unstable circular orbit which it has as both the past and future asymptote, and it thus represents, within orbits of given angular momentum, the boundary between the ``eternal'' ones (bouncing in the potential valley) and those which plunge to the center over the potential maximum defining the circular orbit.

Let us start from the pseudo-Newtonian Schwarzschild black hole as described by the Nowak-Wagoner potential. Using the effective-potential method, we obtain the usual equation for radial velocity\footnote
{Since all our unperturbed fields are spherically symmetric, geodesic motion in them is planar. Conventionally, one adjusts the coordinates so that the orbit under consideration lies in the ``equatorial'' plane ($\theta\!=\!\pi/2$). Then, among others, the angular momentum has just the azimuthal component ($\ell\!\equiv\!\ell_\phi$). In general (without adapting the coordinates in such a manner), one should use $\ell$ instead of $\ell_\phi$, however. We better follow this generic notation in this section, since after a perturbation the field is no longer spherically symmetric and one has to distinguish between $\ell$ and $\ell_\phi$ in any case.} 
\begin{equation}  \label{NW-radialmotion}
  \frac{1}{2}\,(v^r)^2
     = {\cal E}-\left[\frac{\ell^2}{2r^2}-\frac{M}{r}\left(1-\frac{3M}{r}+\frac{12M^2}{r^2}\right)\right]
     =:{\cal E}-V_{\rm eff}(r)\geq 0,
\end{equation}
where $v^r\!=p_r/m$, ${\cal E}\!:=\!E/m$ and $\ell\!:=\!L/m$.
From a smaller $r$-root of equation $\frac{{\rm d}V_{\rm eff}}{{\rm d}r}(r)=0$, one finds the hyperbolic fixed point represented by unstable circular orbit,
\begin{equation}  \label{R0,Newt}
  \ell^2=\frac{M}{r}\left(r^2-6Mr+36M^2\right)
  \quad\Longleftrightarrow\quad
  r=R_0=\frac{\ell^2+6M^2-\sqrt{\ell^4+12M^2\ell^2-108M^4}}{2M} \;.
\end{equation}
The corresponding energy is given by the value of $V_{\rm eff}$ calculated for the above $\ell$ and $r\!=\!R_0$,
\begin{equation}
  {\cal E}_0 = V_{\rm eff}(R_0) = -\frac{M\,(R_0^2-12M^2)}{2R_0^3} \;.
\end{equation}

Homoclinic orbit has the same energy as the hyperbolic-point orbit, therefore, for some chosen value of $\ell$, it is found from equation (\ref{NW-radialmotion}) with ${\cal E}\!=\!{\cal E}_0$ and $R_0$ introduced from above. The orbit has apocentre ($r_{\rm max}$) where $V_{\rm eff}$ again -- at $r\!>\!R_0$ -- reaches the energy level ${\cal E}_0$; see Fig. \ref{homoclinic-orbits}.\footnote
{Note that the homoclinic orbit of course exists for certain range of $\ell$ values only (we will specify this range later).}
We will derive explicit solution from the equation for radial motion (\ref{NW-radialmotion}) rewritten in terms of the reciprocal radius $u:=\frac{1}{r}\,$: expressing
\begin{equation}
  v^r\equiv\frac{{\rm d}r}{{\rm d}t}
     = \frac{{\rm d}r}{{\rm d}u} \frac{{\rm d}u}{{\rm d}\vartheta} \frac{{\rm d}\vartheta}{{\rm d}t}
     = -\ell\,\frac{{\rm d}u}{{\rm d}\vartheta} \;,
\end{equation}
it takes the form
\begin{equation}
  \left(\frac{{\rm d}u}{{\rm d}\vartheta}\right)^{\!2}
     = \frac{2{\cal E}+2Mu(12M^2 u^2-3Mu+1)-u^2\ell^2}{\ell^2} \;.
\end{equation}
The latter is specified to the desired homoclinic orbit by substituting ${\cal E}\!=\!{\cal E}_0$; it can then be further rewritten as
\begin{equation}
  \left(\frac{{\rm d}u}{{\rm d}\vartheta}\right)^{\!2}
     = \frac{24M^3}{\ell^2}\,(u-U_0)^2(u-u_{\rm max}),
  \qquad
  u_{\rm max}=\frac{\ell^2+6M^2}{24M^3}-2U_0
             =U_0-\frac{\sqrt{\ell^4+12M^2\ell^2-108M^4}}{24M^3} \;,
\end{equation}
where $U_0\!:=\!1/R_0$ and $u_{\rm max}\!:=\!1/r_{\rm max}$ (needless to say, $u_{\rm max}$ is actually a {\em minimum} of $u$).
Solving the last equation for $u(\vartheta)$, one obtains the homoclinic orbit,
\begin{equation}  \label{u,homoclinic,NW}
  u_0(\vartheta)=u_{\rm max}+
                 (U_0-u_{\rm max})\tanh^2\!\left(\frac{M}{\ell}\,\vartheta\,\sqrt{6M(U_0-u_{\rm max})}\right).
\end{equation}
Note that it is suitably parametrized: $u_0(0)\!=\!u_{\rm max}$ (the turning point) and $u_0(\vartheta\!\to\!\pm\infty)=U_0$. The orbit is shown in Fig. \ref{homoclinic-orbits} in polar coordinates $r\!=\!\frac{1}{u}$ and $\vartheta$ (it is planar, lying in the equatorial plane of the configuration space).

Now to the relativistic case. For the Schwarzschild black hole, one has the well-known radial equation
\begin{equation}  \label{Schw-radialmotion}
  \left(\frac{{\rm d}u}{{\rm d}\vartheta}\right)^{\!2}
    =\frac{{\cal E}^2-V_{\rm eff}^2}{\ell^2} \;,
  \quad\qquad
  V_{\rm eff}^2=(1-2Mu)(1+\ell^2 u^2)
\end{equation}
and from there the hyperbolic-orbit parameters
\begin{equation}  \label{R0,Schw}
  \ell^2=\frac{Mr^2}{r-3M}
  \quad \Longleftrightarrow \quad
  r=R_0=\frac{\ell}{2M}\left(\ell-\sqrt{\ell^2-12M^2}\right);
  \quad\qquad
  {\cal E}_0^2 = \frac{(R_0-2M)^2}{R_0(R_0-3M)} \;.
\end{equation}
Fixing the energy to that of the unstable circular orbit, ${\cal E}\!=\!{\cal E}_0$, the radial equation (\ref{Schw-radialmotion}) can be rewritten as
\begin{equation}
  \left(\frac{{\rm d}u}{{\rm d}\vartheta}\right)^{\!2}
     =2M\,(u-U_0)^2(u-u_{\rm max}) \,,
  \quad\qquad
  u_{\rm max}=\frac{1}{2M}-2U_0
             =U_0-\frac{\sqrt{\ell^2-12M^2}}{2M\ell} \;,
\end{equation}
which yields a solution similar to (\ref{u,homoclinic,NW}) for the homoclinic orbit, just with different coefficients:
\begin{equation}  \label{u,homoclinic,Schw}
  u_0(\vartheta)=u_{\rm max}+
                 (U_0-u_{\rm max})\tanh^2\!\left(\frac{1}{2}\,\vartheta\,\sqrt{2M(U_0-u_{\rm max})}\right).
\end{equation}
This result was already obtained by \cite{BombelliC-92} (just with the reciprocal radius chosen as $x\!:=\!2M/r$) in their study of chaos in the Schwarzschild background periodically perturbed by gravitational waves.\footnote
{In that case, the Melnikov method can be used in its original form. \cite{BombelliC-92} are also recommended for a thorough introduction to the method.}

For the extreme Reissner-Nordstr\"om black hole, the radial equation reads
\begin{equation}  \label{RN-radialmotion}
  \left(\frac{{\rm d}u}{{\rm d}\vartheta}\right)^{\!2}
    =\frac{{\cal E}^2-V_{\rm eff}^2}{\ell^2} \;,
  \quad\qquad
  V_{\rm eff}^2=(1-Mu)^2(1+\ell^2 u^2)
\end{equation}
and yields, for the hyperbolic circular orbit,
\begin{equation}  \label{R0,RN}
  \ell^2=\frac{Mr^2}{r-2M}
  \quad \Longleftrightarrow \quad
  r=R_0=\frac{\ell}{2M}\left(\ell-\sqrt{\ell^2-8M^2}\right);
  \quad\qquad
  {\cal E}_0^2 = \frac{(R_0-M)^3}{R_0^2(R_0-2M)} \;.
\end{equation}
In contrast to the Schwarzschild case (\ref{Schw-radialmotion}), the right-hand side of (\ref{RN-radialmotion}) is a polynomial of the fourth order (in $u$) which has {\em three} stationary points: one is the desired unstable periodic orbit, another one is the usual stable circular orbit, and the third one is located at the horizon $r\!=\!M$; there, $V_{\rm eff}\!=\!0$ and it is a minimum (the ``circular orbit'' on the horizon however has $p_t\!=\!0$, which implies $p_i\!=\!0$, so it actually corresponds to the light-like horizon generator).
Like in the Schwarzschild case, we denote $U_0\!\equiv\!1/R_0$, fix the energy by ${\cal E}_0\!=\!V_{\rm eff}(R_0)$ and rewrite the equation for the homoclinic orbit as
\begin{equation}  \label{RN,eq-for-homoclinic}
  \left(\frac{{\rm d}u}{{\rm d}\vartheta}\right)^{\!2}
     =M^2(u-U_0)^2(u-u_{\rm max})(u_{\rm min}-u) \,,
  \quad\qquad
  u_{\rm max/min}=\frac{1}{M}-U_0\mp\sqrt{\frac{U_0}{M}} \;,
\end{equation}
$u_{\rm min}$ denoting a ``pericentre'' now located inside the black hole. We thus conclude that the homoclinic orbit is located in the interval $u\in (u_{\rm max},U_0)$ or, equivalently, $r\in (U_0,r_{\rm min})$ (bear in mind that $u_{\rm max}\leq U_0\leq u_{\rm min}$). Solving equation (\ref{RN,eq-for-homoclinic}) with the condition that $u_{\rm max}$ corresponds to the outer turning point of the homoclinic orbit, we have
\begin{equation}  \label{u,homoclinic,RN}
  u_0(\vartheta)
     =U_0+\frac{2\,(u_{\rm min}-U_0)(U_0-u_{\rm max})}
               {2U_0-u_{\rm max}-u_{\rm min}
                -(u_{\rm min}-u_{\rm max})\cosh\!\left(M\vartheta\,\sqrt{(u_{\rm min}-U_0)(U_0-u_{\rm max})}\right)}
  \;.
\end{equation}
We can again see that $u_0(\vartheta\to\pm\infty)=U_0$. The spatial representation (polar graph) of this homoclinic orbit is shown in  Fig. \ref{homoclinic-orbits}.

\section{Calculation of the Melnikov function}
\label{Melnikov}

Consider first that the unperturbed angular velocity $\dot\psi$ has the same form in the pseudo-Newtonian as well as in the relativistic case,
\begin{equation}
  \dot\psi(r,J)=\frac{\partial H_0}{\partial J_\vartheta}
               =\frac{J_\vartheta+L_\phi}{mr^2}
               =\frac{\ell}{r^2} \;.
\end{equation}

For the pseudo-Newtonian Hamiltonian (\ref{H,Newt})--(\ref{H1,Newt}), with the exterior potential $\nu_{\rm ext}$ transformed to the spheroidal coordinates according to (\ref{rho,z-Weyl}), the Poisson brackets yield
\begin{equation}
  \left\{H_0,\frac{H_1}{\dot\psi}\right\}
    =\frac{\partial H_0}{\partial r}\,\frac{\partial}{\partial p_r}\left(\frac{H_1}{\dot\psi}\right)
     -\frac{\partial H_0}{\partial p_r}\,\frac{\partial}{\partial r}\left(\frac{H_1}{\dot\psi}\right)
    =-\frac{p_r}{m}\left(\frac{r^2}{\ell}\frac{\partial H_1}{\partial r}+2H_1\frac{r}{\ell}\right)
\end{equation}
and so the complete integrand of the Melnikov function (\ref{Melnikov,new}) reads
\begin{equation}
  \frac{1}{\dot\psi}\left\{H_0,\frac{H_1}{\dot\psi}\right\}
    =-v^r \frac{r^3}{\ell^2}\left(2H_1+r\,\frac{\partial H_1}{\partial r}\right).
\end{equation}
Performing the canonical transformation $H_1(r,\theta)\to H_1(r,L,\vartheta,L_\phi)$ according to (\ref{canonical-trans}), inserting the homoclinic orbit (\ref{u,homoclinic,NW}) and the corresponding radial velocity as
\[r_0(\vartheta)=\frac{1}{u_0(\vartheta)} \,,
  \qquad
  v_0^r(\vartheta)=-\ell\,\frac{{\rm d}u_0(\vartheta)}{{\rm d}\vartheta}\,,\]
and integration along the homoclinic orbit yields
\begin{equation}  \label{Melnikov,final,Newt}
  M(\vartheta_0)
     =-\int\limits_{-\infty}^{\infty}v_0^r(\vartheta)\,\frac{r_0^3(\vartheta)}{\ell^2}
       \left[2H_1(r_0(\vartheta),\vartheta+\vartheta_0)
             +r_0(\vartheta)\,\frac{\partial H_1}{\partial r}(r_0(\vartheta),\vartheta+\vartheta_0)\right]
       {\rm d}\vartheta \,.
\end{equation}
Note that the resulting Melnikov function $M(\vartheta_0)$ depends, besides $\vartheta_0$, on the fixed parameters $L$ and $L_\phi$ as well, but we do not indicate this.
 
For the relativistic Hamiltonian (\ref{H,rel}) with (\ref{metric,expansion}), one has the perturbation
\begin{equation}
  H_1=-\frac{{\cal M}}{2m}
       \left[\frac{\frac{\partial g_{tt}}{\partial {\cal M}}}{(g_{tt})^2}\,E^2
             +\frac{\frac{\partial g_{rr}}{\partial {\cal M}}}{(g_{rr})^2}\,p_r^2
             +\frac{\frac{\partial g_{\theta\theta}}{\partial {\cal M}}}{(g_{\theta\theta})^2}
              \left(L^2-\frac{L_\phi^2}{\sin^2\theta}\right)
             +\frac{\frac{\partial g_{\phi\phi}}{\partial {\cal M}}}{(g_{\phi\phi})^2}\,L_\phi^2\right],
\end{equation}
where all the metric terms are evaluated at ${\cal M}\!=\!0$.
There, one first transforms the metric due to the exterior sources, usually presented in the Weyl coordinates, to the spheroidal (Schwarzschild-type) ones, which means to use relations (\ref{rho,z-Weyl}) for the Bach-Weyl ring or the thin disc, while the ``extreme-type'' relations (\ref{rho,z-Weyl,extreme}) for the Majumdar-Papapetrou ring. Then one performs the canonical transformation (\ref{canonical-trans}).

Before embarking on the Melnikov function, recall that it provides the linear-in-perturbation part of the distance between the asymptotic manifolds, i.e., the whole method works in the ${\cal O}({\cal M}/M)$ order. Consequently, it is sufficient to use the unperturbed metric when raising and lowering indices of quantities which appear inside the Melnikov function. In particular, we have $p_r=g_{rr}({\cal M}\!=\!0)\,p^r+{\cal O}({\cal M})$, and the term containing $\frac{\partial g_{rr}}{\partial {\cal M}}$ would contribute as ${\cal O}({\cal M}^2)$, so it is possible to neglect it. (Let us stress once more that this neglection only concerns raising and lowering of the indices.)

Now, the Poisson brackets taken in $(r,p_r)$ read
\begin{align}
  \left\{H_0,\frac{H_1}{\dot\psi}\right\}
    &=\frac{\partial H_0}{\partial r}\,
      \frac{{\cal M}}{\dot\psi}\,\frac{\partial g^{rr}}{\partial {\cal M}}\,\frac{p_r}{m}
      -\frac{g^{rr}p_r}{m}\left(\frac{1}{\dot\psi}\frac{\partial H_1}{\partial r}+\frac{2r}{\ell} H_1\right)=
      \nonumber \\
    &=-u^r\frac{r}{\ell}
      \left(r\,\frac{\partial H_0}{\partial r}\,
            \frac{{\cal M}\frac{\partial g_{rr}}{\partial {\cal M}}}{g_{rr}}
            +r\,\frac{\partial H_1}{\partial r}+2H_1\right),
\end{align}
so, substituting $\dot\psi\!=\!\ell/r^2$, the Melnikov integral is
\begin{equation}  \label{Melnikov,final,rel}
  M(\vartheta_0)
     =-\int\limits_{-\infty}^{\infty}
       \left[u^r\frac{r^3}{\ell^2}\left(r\,\frac{\partial H_0}{\partial r}\,
             \frac{{\cal M}\frac{\partial g_{rr}}{\partial {\cal M}}}{g_{rr}}
             +2H_1+r\,\frac{\partial H_1}{\partial r}\right)\right]
       \left(r_0(\vartheta),u_0^r(\vartheta),\vartheta+\vartheta_0\right)\,{\rm d}\vartheta \;,
\end{equation}
where the metric terms are again evaluated at ${\cal M}\!=\!0$, the dependence on the fixed parameters $L$ and $L_\phi$ is not explicitly indicated, and, like in the previous case, the homoclinic orbit is expressed in terms of $u_0(\vartheta)$, where
\[r_0(\vartheta)=\frac{1}{u_0(\vartheta)} \,,
  \qquad
  u_0^r(\vartheta)=\frac{{\rm d}r_0(\vartheta)}{{\rm d}\tau}=-\ell\,\frac{{\rm d}u_0(\vartheta)}{{\rm d}\vartheta} \,,
  \qquad
  p_r=m g_{rr}({\cal M}\!=\!0)\,u^r \,.\]
 
The Melnikov integrals (\ref{Melnikov,final,Newt}) and (\ref{Melnikov,final,rel}) can now be evaluated numerically. We shall see that they indeed have simple zero points.

%%%%%%%%%%%%%%%%%%%%%%%%%%%%%%%%%%%%
\begin{figure}[h!]\centering
\includegraphics[width=\textwidth]{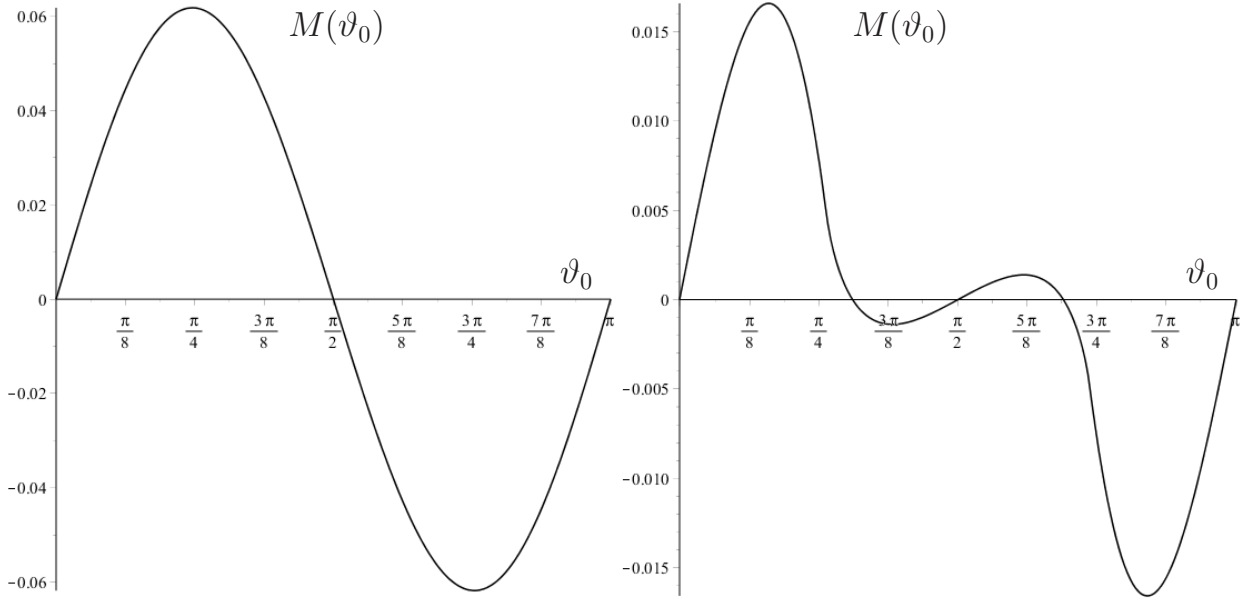}
\caption{Examples of the Melnikov-function behaviour. Left plot: $M(\vartheta_0)$ for the extreme Reissner-Nordstr\"om black hole encircled by a light MP ring at $b\!=\!15M$, for orbital parameters ${\cal E}\!=\!0.9428$, $\ell\!=\!3M$, $\ell_\phi\!=\!0$. Right plot: $M(\vartheta_0)$ for the pseudo-Newtonian (Nowak-Wagoner) Schwarzschild-type potential encircled by a light inverted first MM disc at $b\!=\!6M$, for orbital parameters ${\cal E}\!=\!-0.0510$, $\ell\!=\!2.5M$, $\ell_\phi\!=\!M$. The sine-like shape is in general more frequent than the more complicated behaviour shown in the right plot.}
\label{Melnikov-examples}
\end{figure}

\begin{figure}[h!]\centering
\includegraphics[width=\textwidth]{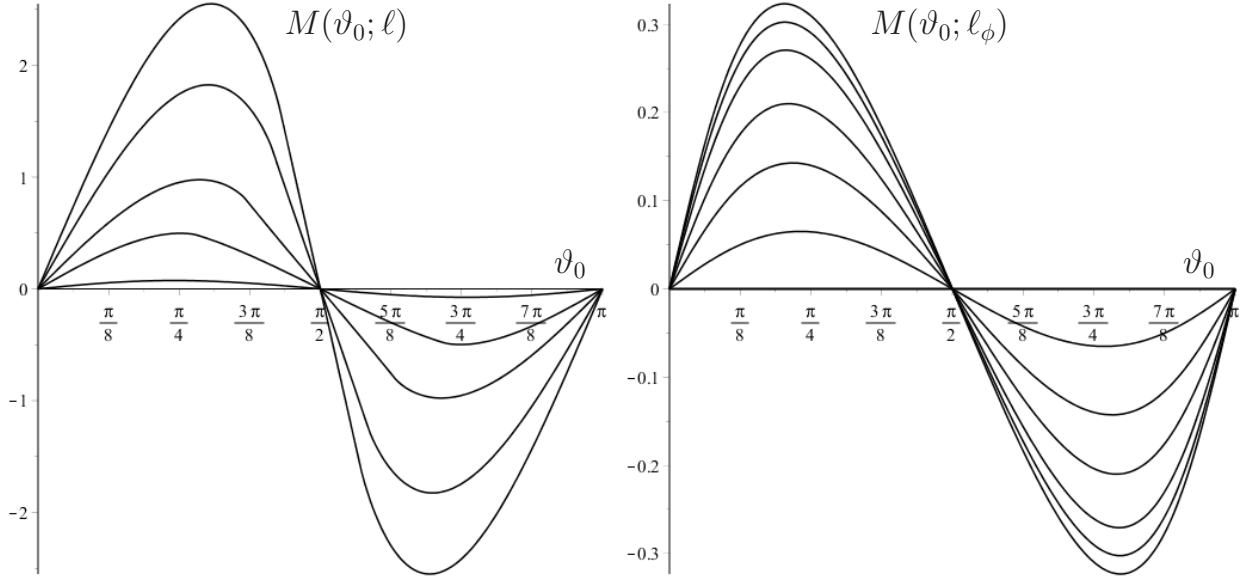}
\caption{Dependence of the Melnikov function on angular momentum, illustrated on the pseudo-Newtonian (Nowak-Wagoner) Schwarzschild black hole surrounded by a light inverted first MM disc. Left plot: dependence of $M(\vartheta_0)$ on total angular momentum $\ell$ for a disc at $b\!=\!10M$ and for $\ell_\phi\!=\!M$. The wave-shape amplitude grows with increasing $\ell$; the shown curves correspond to $\ell\!=\!2.5M$, $2.55M$, $2.6M$, $2.7M$ and $2.8M$ (the respective orbital energies are ${\cal E}\!=\!-0.05098$, $-0.04521$, $-0.03839$, $-0.02167$, $-0.00068$). Right plot: dependence of $M(\vartheta_0)$ on $\ell_\phi$ for a disc at $b\!=\!20M$, for $\ell\!=\!2.6M$ and ${\cal E}\!=\!-0.03839$. The wave-shape amplitude increases with decreasing $\ell_\phi$ (and fixed $\ell$); the shown curves correspond to $\ell_\phi\!=\!2.6M$ (this yields $M(\vartheta_0)\!=\!0$), $2.5M$, $2.3M$, $2M$, $1.5M$, $1M$ and $0$.}
\label{Melnikov-l-lphi}
\end{figure}

\begin{figure}[h!]\centering
\includegraphics[width=\textwidth]{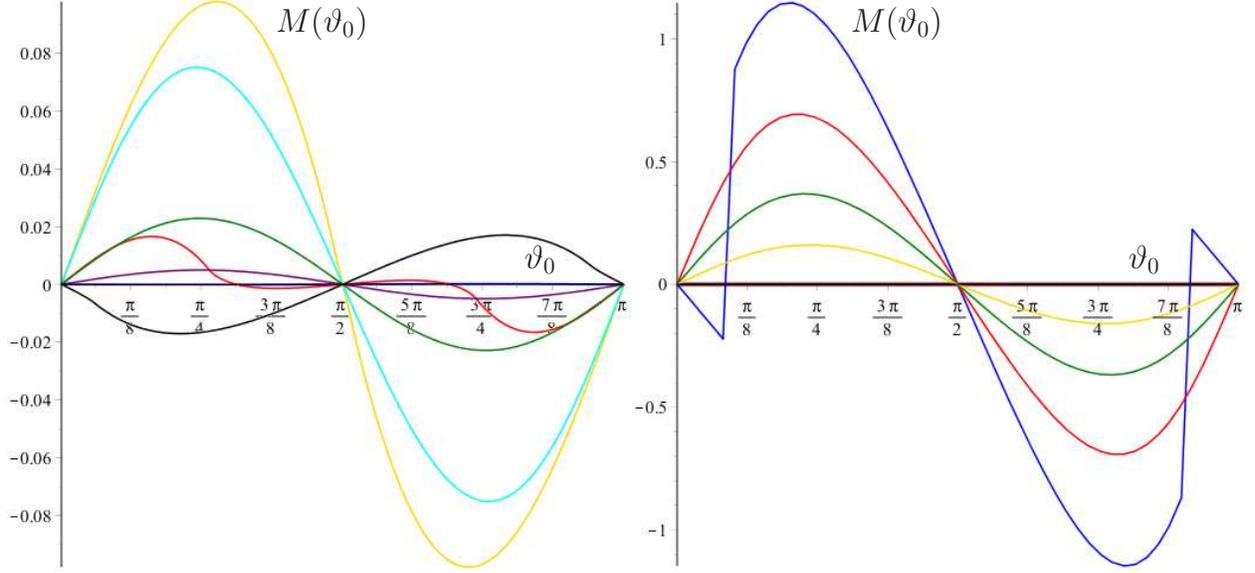}
\caption{Dependence of the Melnikov function on the external-source Weyl radius $b$. Left plot: for the pseudo-Newtonian (Nowak-Wagoner) Schwarzschild black hole surrounded by a light inverted first MM disc, with orbital parameters ${\cal E}\!=\!-0.05098$, $\ell\!=\!2.5M$, $\ell_\phi\!=\!M$. Line colouring: $b\!=\!M$ blue (almost coincides with the axis), $b\!=\!5M$ black, $b\!=\!6M$ red, $b\!=\!7M$ yellow, $b\!=\!10M$ light blue, $b\!=\!15M$ green, $b\!=\!25M$ violet. Right plot: similar plot for the extreme Reissner-Nordstr\"om black hole surrounded by a light MP ring, with orbital parameters ${\cal E}\!=\!0.9428$, $\ell\!=\!3M$, $\ell_\phi\!=\!M$. Line colouring: $b\!=\!0$ brown (coincides with the axis), $b\!=\!10M$ blue, $b\!=\!12M$ red, $b\!=\!15M$ green, $b\!=\!20M$ yellow. (Note that $b\!=\!0$ means that the MP-ring Weyl radius is the same as that of the horizon, which for the extreme RN black hole is a viable option.)}
\label{Melnikov-b}
\end{figure}

\begin{figure}[h!]\centering
\includegraphics[width=\textwidth]{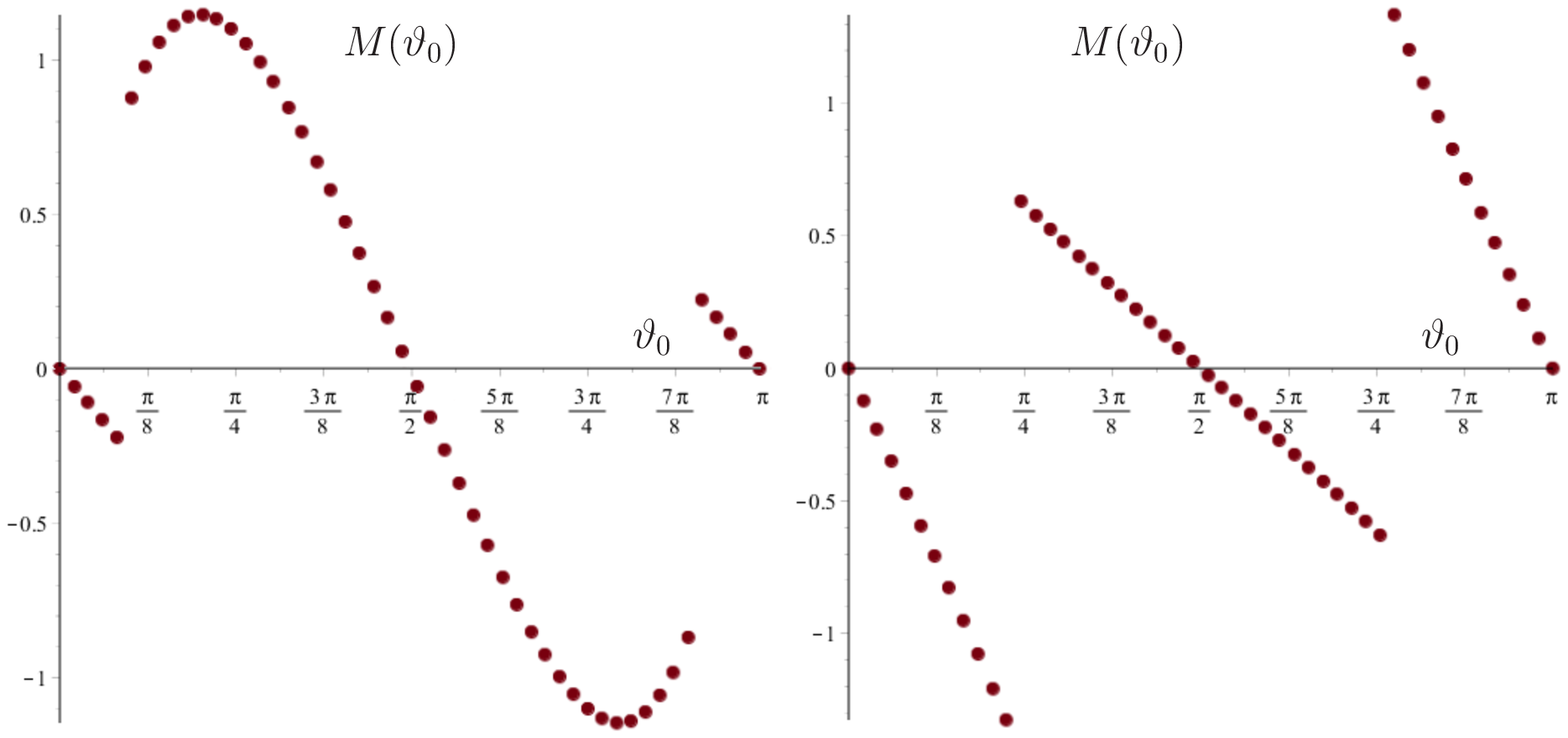}
\caption{Two examples of a discontinuity of the Melnikov function which may appear at the location of the external source. They have been obtained for two different configurations of the extreme Reissner-Nordstr\"om black hole encircled by the MP ring. (See the main text for a comment on this feature.)}
\label{Melnikov-discont}
\end{figure}

%%%%%%%%%%%%%%%%%%%%%%%%%%%%%%%%%%%%

\section{Numerical evaluation of the Melnikov function and comparison with Poincar\'e diagrams}
\label{numerics}

\subsection{Behaviour of the Melnikov function}

The above-derived Melnikov functions can now be evaluated numerically on the interval of their periodicity, i.e. $(0,\pi)$. Though they are rather complicated and cannot be expressed in closed form, their plots are surprisingly simple, namely, they mostly exhibit a sine-like behaviour as can be seen in Fig. \ref{Melnikov-examples}: the plot (a) shows a typical behaviour, while the shape shown in plot (b) is much less frequent. In any case, our experience is that, for our ring or disc perturbations, the Melnikov function $M(\vartheta_0)$ typically does have simple zeros.

However, let us look first at how the Melnikov function depends on free parameters. These are three -- the total angular momentum $\ell$, the $\phi$-component of the angular momentum, $\ell_\phi$ (both on unit particle rest mass), and ring/disc radius $b$. The angular momentum $\ell$ selects one particular homoclinic orbit, but it is not an integral of motion of the complete system. (The homoclinic orbit is also fixed uniquely by energy ${\cal E}$ or by $J_\vartheta$, but we will use $\ell$.)

The dependence on $\ell$ is very simple: the amplitude of $M(\vartheta_0)$ just increases with $\ell$ ranging within its possible interval (see below, Section \ref{ranges}), as illustrated in Fig. \ref{Melnikov-l-lphi} (left plot).

The effect of $\ell_\phi$, with $b$ and $\ell$ fixed, is (also) the same for all the source configurations we consider: $M(\vartheta_0)$ has a maximal amplitude for $\ell_\phi\!=\!0$ and goes to zero when $\ell_\phi$ approaches $\ell$, $\lim\limits_{\ell_\phi\to\ell} M(\vartheta_0)\!=\!0$ (since $\ell_\phi$ is a component of the angular-momentum vector whose norm is $\ell$, we have $\ell\geq\!\ell_\phi\!\geq\!0$, where, without loss of generality, we consider $\ell_\phi\geq\!0$). This is an expectable behaviour: for $\ell_\phi\!=\!\ell$ the motion is confined to the equatorial plane fixed by the disc/ring; this plane, however, is the plane of reflectional symmetry and the complete system is independent of $\vartheta$, so the homoclinic orbit is preserved in that case (asymptotic manifolds stay coinciding along it).
The dependences of Melnikov function on $\ell$ and on $\ell_\phi$ are exemplified in Fig. \ref{Melnikov-l-lphi} for the pseudo-Newtonian Schwarzschild-type field (for the relativistic fields we found very similar results).

The last free parameter is the ring/disc radius $b$. For large increasing $b$, the amplitude of $M(\vartheta_0)$ gradually decreases to zero, because the perturbation $H_1$ vanishes, $\lim\limits_{b\to \infty}H_1=0$ (a ring/disc which is infinitely far is not being felt).
In the opposite limit $b\to 0$, the result depends on the gravitational perturbation. The Majumdar-Papapetrou ring behaves, in this limit, as the extreme Reissner-Nordstr\"om black hole (see previous paper \citealt{PolcarSS-19}), which again leads to $\lim\limits_{b\to 0}M(\vartheta_0)=0$. The potentials $\nu_{\rm ext}(\rho,z)$ of the Bach-Weyl ring and of the inverted first Morgan-Morgan disc have the same limit,
\[\lim\limits_{b\to 0} \nu_{_{\rm BW,iMM}}(\rho,z)=-\frac{\cal M}{\sqrt{\rho^2+z^2}} \;,\]
which corresponds to a point particle located at the origin. One would thus expect preservation of the spherical symmetry and, consequently, vanishing of the Melnikov function. However, this is only the case if we perform the transformation between cylindrical and spheroidal coordinates using the Euclidean relations $\rho\!=\!r\sin\theta$, $z\!=\!r\cos\theta$; if we use instead the ``relativistic'' relations (\ref{rho,z-Weyl}), we get a non-zero $M(\vartheta_0)$ with typical sine-like behaviour. (Note that the relativistic counterpart of the above limit is the Curzon-Chazy metric which really does not represent a simple, monopole particle.) We thus see that choosing the right coordinate transformation can be rather tricky. In this particular situation, we interpret the sources in a Newtonian fashion which means that we assume the space to be flat, and so the Euclidean transformation relation appears to be more adequate. Finally, the dependence of the Melnikov function on radius $b$ can be seen in Fig. \ref{Melnikov-b}.

In Fig. \ref{Melnikov-b}, the blue curve in the right-hand side plot clearly has discontinuity at two values of $\vartheta_0$ symmetrically placed with respect to $\pi/2$. The Melnikov function may really be discontinuous, because the integration path (i.e. the unperturbed homoclinic orbit) may intersect the source. Actually, the source lies at $\theta\!=\!\pi/2$ (which corresponds to $\vartheta\!=\!0$, except for the $L_\phi\!=\!L$ case), while the homoclinic orbit goes over the whole range of $\vartheta$, namely $-\infty\!<\!\vartheta\!<\!+\infty$, which, according to (\ref{canonical-trans}), corresponds to $\theta$ ranging from $\theta_{\rm min}\!=\!\arcsin\frac{L_\phi}{L}\,(\leq\!\pi/2)$ to $\theta_{\rm max}\!=\!\pi\!-\!\theta_{\rm min}\,(\geq\!\pi/2)$ (see section \ref{modification}). Therefore, the homoclinic orbit necessarily intersects the plane of the external source somewhere (if not entirely lying in it, which is the case when $L_\phi\!=\!L$). If the source lies (or extends) below the apocentre of the homoclinic orbit, the integration may thus cross it. Our sources are thin (ring or disc), so they are singular locations of the corresponding perturbation (at least at a certain level of metric derivative), and thus the Melnikov function $M(\vartheta_0)$ in such a case becomes discontinuous for some values of $\vartheta_0$. However, it may still have simple zeros and be continuous in their neighbourhoods (Fig. \ref{Melnikov-discont}), so the homoclinic chaos should appear then as well. This point would anyway be worth further study.

\subsection{How probable is the chaotic regime?}
\label{ranges}

The Melnikov-function simple zeros imply homoclinic chaos, and we have seen that for our systems the homoclinic orbits really ``break up'', but a different question is how large the parameter space is for which the homoclinic orbits at all exist.
Looking first at the relations (\ref{R0,Newt}), (\ref{R0,Schw}) and (\ref{R0,RN}) for unstable circular orbits (hyperbolic fixed points) $R_0$, we see they all contain square roots of arguments which are only non-negative for a sufficiently large $\ell$. Another condition for the existence of the homoclinic orbit is that there has to exist, for the value of energy fixed by the unstable circular orbit, a turning point at larger radius (this corresponds to apocentre of the homoclinic orbit, $r_{\rm max}$). For our effective potentials this condition is satisfied if
\[V_{\rm eff}(R_0)\leq \lim\limits_{r\to \infty} V_{\rm eff}(r).\]
This leads, on the contrary, to an {\em upper} bound for $\ell$, because the potential maximum corresponding to the unstable circular orbit must not be too high. Thus obtained conditions $\ell_{\rm min}\!<\ell\leq\ell_{\rm max}$ can be translated in conditions for the integrals of motion ${\cal E}$ and $\ell_\phi$. 

The summary of necessary conditions is actually clear from the effective-potential behaviour, as seen in Fig. \ref{homoclinic-orbits}: the unperturbed potential must have a maximum between the horizon and radial infinity, with lower value than the potential reaches asymptotically.
By the particle energy ${\cal E}$, the geodesics in the unperturbed system can be divided into three groups, independently of their angular momentum:
\begin{enumerate}
  \item ${\cal E}\!<\!{\cal E}_{\rm min}\!:=\!{\cal E}(\ell\!=\!\ell_{\rm min})$:
  geodesics which always end in the black hole and cannot reach infinity (or come from there).
  \item ${\cal E}\in ({\cal E}_{\rm min},{\cal E}_{\rm max}\rangle$:
  geodesics which end in the black hole plus eternal bound orbits (these also cannot exist at asymptotic radii).
  \item ${\cal E}\!>\!{\cal E}_{\rm max}\!:=\!{\cal E}(\ell\!=\!\ell_{\rm max})\!=\!V_{\rm eff}(r\!\to\!\infty)$:
  geodesics which can exist at infinite $r$ as well as end in the black hole; they may be reflected by the centrifugal barrier from both sides. 
\end{enumerate}
We can expect that the same behaviour occurs in the perturbed system provided that the perturbation is small enough. As expected, only the second of the above cases leads to chaotic dynamics since only in that case there are bound orbits separated from the ingoing/outgoing orbits by a separatrix.

In terms of $\ell_\phi$\,, the upper bound naturally reads $\ell_\phi\!\leq\!\ell_{\rm max}$\,, while the lower bound may not exist, because $\ell_\phi\!<\!\ell_{\rm min}$ does not mean that $\ell\!\leq\!\ell_{\rm min}$ ($\ell$ also has the other component).
We thus rather specify the intervals for which the homoclinic orbit does exist in terms of $\ell$ and ${\cal E}$ -- see Table 1.
To summarize, the Melnikov method implies, for our black-hole systems, that on any hypersurface given by ${\cal E}\in ({\cal E}_{\rm min},{\cal E}_{\rm max}\rangle$ and $\ell_\phi\in\langle 0,\ell_{\rm max})$, there exist transverse homoclinic orbits in the neighbourhood of which the system exhibits chaotic behaviour.

\begin{center}
\begin{tabular}{ |c|c|c|c| } 
\hline
 \multicolumn{4}{|c|}{Table 1: Conditions for the existence of homoclinic orbits} \\
  \hline
  \small potential &\small Nowak-Wagoner pseudo-Newtonian &\small Schwarzschild &\small extreme Reissner-Nordstr\"om \\
  \hline 
  \small interval of $\ell$ & $\scriptstyle \left(\sqrt{6},\,\sqrt{8\sqrt{3}-6}\right)M\,\doteq\,(2.45,2.80)M$
                            & $\scriptstyle \left(2\sqrt{3},\,4\right)M\,\doteq\,(3.46,4.00)M$
                            & $\scriptstyle \left(2\sqrt{2},\,\frac{1}{2}\sqrt{22+10\,\sqrt{5}}\right)M\,\doteq\,(2.83,3.33)M$ \\
  \hline
  \small interval of ${\cal E}$ & $\scriptstyle \left(-\frac{1}{18},\,0\right)$
                                & $\scriptstyle \left(\frac{2}{3}\sqrt{2},\,1\right)\,\doteq\,(0.94,1.00)M$
                                & $\scriptstyle \left(\frac{3}{8}\sqrt{6},\,1\right)\,\doteq\,(0.92,1.00)M$\\
  \hline
\end{tabular}
\end{center}

\subsection{Comparison with numerical geodesic dynamics}

Although there is no reasonable doubt that the Melnikov method really works, it is always interesting to compare analytic ``predictions'' with computation of an actual flow of a given system from particular initial conditions. We will do so, numerically, for the specific case of the extreme Reissner-Nordstr\"om black hole encircled by the Majumdar-Papapetrou ring, on equatorial Poincar\'e diagrams drawn in the $(r,u^r)$ axes. We fix the values of the basic integrals of motion as ${\cal E}\!=\!0.942809$ and $\ell\!=\!M$; these values indeed fall within the intervals given in Table 1, so one can expect that the numerical dynamics will involve chaotic layers spreading from separatrices of the unperturbed system.

%%%%%%%%%%%%%%%%%%%%%%%%%%%%%%%%%%%%

\begin{figure}[h!]\centering
\includegraphics[width=\textwidth]{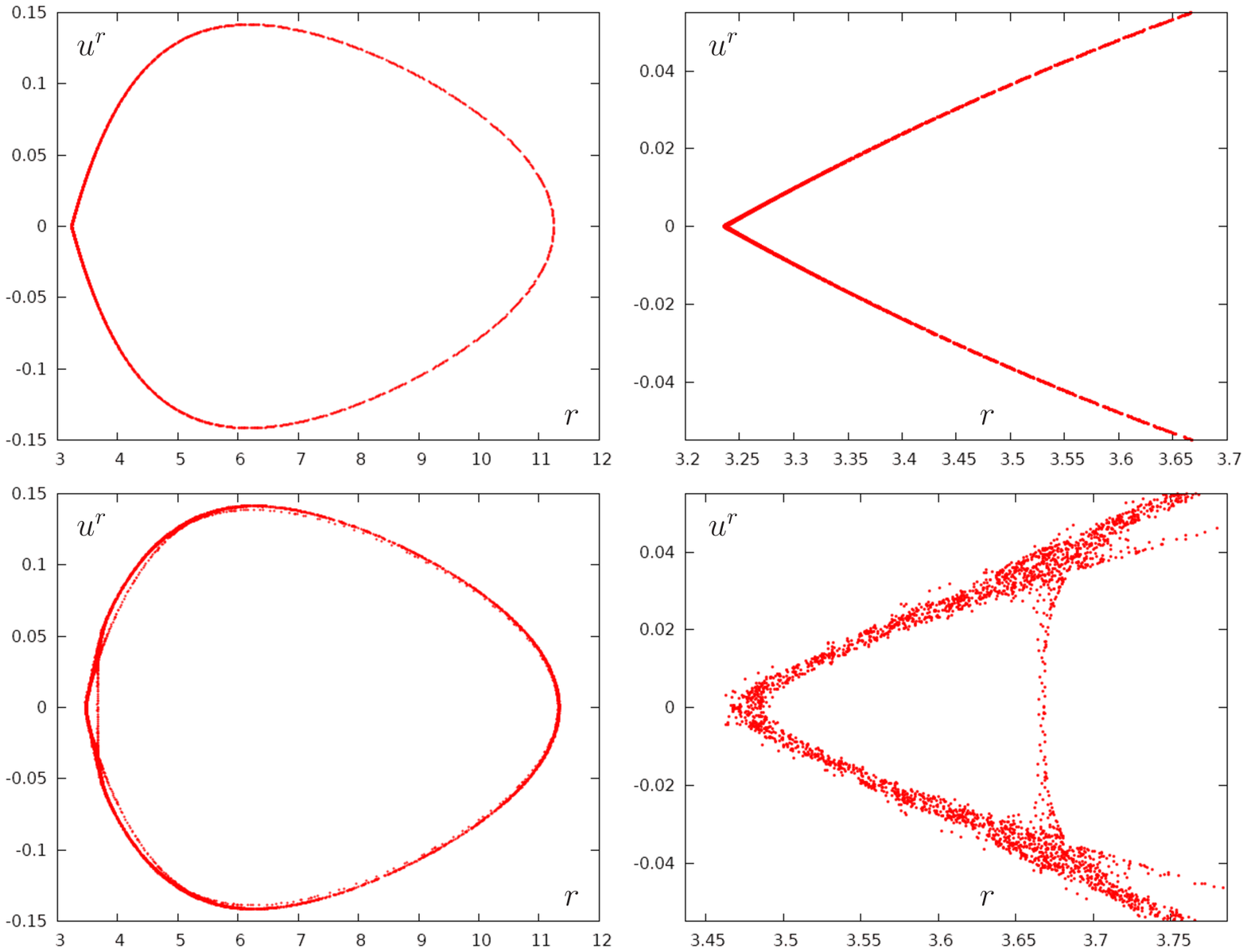}
\caption{Equatorial Poincar\'e section of a typical orbit lying close to a separatrix of an extreme Reissner-Nordstr\"om black hole (of mass $M$), before (top row) and after (bottom row) a perturbation due to an MP ring with mass ${\cal M}\!=\!0.01M$. The orbit has constants of motion ${\cal E}\!=\!0.942809$, $\ell\!=\!M$. The left column shows the entire orbit, while the right column shows a detail of a vicinity of the corresponding unstable circular orbit.}
\label{breakup}
\end{figure}

\begin{figure}[h!]\centering
\includegraphics[width=\textwidth]{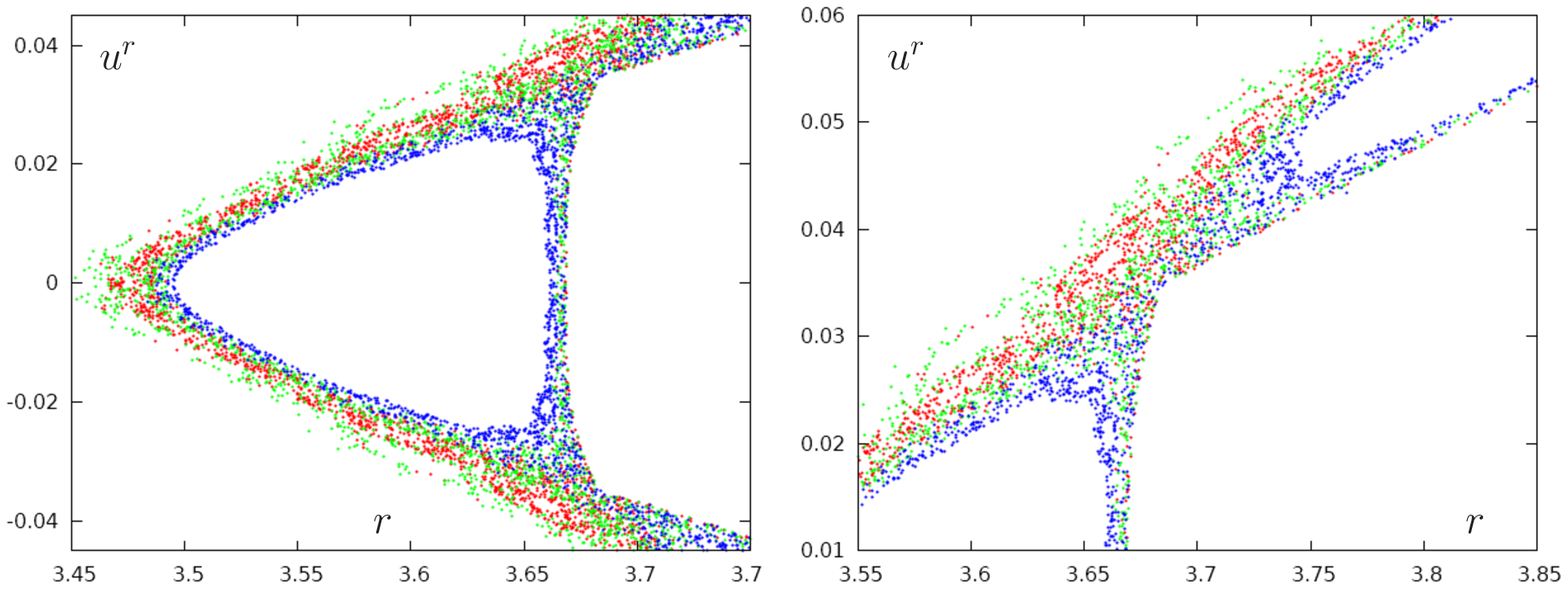}
\caption{A detailed zoom (even more detailed on the right-hand side) of the equatorial Poincar\'e diagram of three different orbits (having ${\cal E}\!=\!0.942809$, $\ell\!=\!M$) passing close to the unstable periodic orbit of the system of extreme RN black hole perturbed by the MP ring with ${\cal M}\!=\!0.01M$. The orbits are distinguished by colours.} 
\label{breakup-detail}
\end{figure}

%%%%%%%%%%%%%%%%%%%%%%%%%%%%%%%%%%%%

We start with the unperturbed system containing only the extreme Reissner-Nordstr\"om black hole (the ring mass is set to zero). We select an orbit close to the separatrix whose Poincar\'e section is seen in the top row of Fig. \ref{breakup}; it is a smooth curve as expected for a regular orbit (its detail close to the hyperbolic circular orbit is added in the right plot of the top column). 
After adding a Majumdar-Papapetrou ring with mass ${\cal M}\!=\!0.01M$, the Poincar\'e section changes to that given in the bottom row of Fig. \ref{breakup} (the neighbourhood of the hyperbolic circular orbit is again magnified in the right plot). Fig. \ref{breakup-detail} shows three more similar geodesics obtained for the perturbed field. Obviously, they all densely fill a certain area around the separatrix, confirming their chaotic nature. On the other hand, farther from the separatrix, there is no sign of chaotic dynamics. Actually, a typical Poincar\'e section of a situation with accessible region open towards the black hole contains, after weak perturbation, a central regular island surrounded by sparse traces of those orbits which plunge into the black hole, and with only a thin chaotic layer arisen from the separatrix lying between the bound orbits and the plunging orbits. This is in fact a common experience -- for instance, a very similar result was obtained by \cite{Aslanov-16} when checking numerically the Melnikov-method predictions for quite a different system, namely a tow of space debris by a tether. It also appears that the chaos which develops in our systems is {\em exclusively} of the homoclinic origin.

\subsection{Remarks on the Melnikov method}

Let us stress again that the suggested canonical transformation brings our problem to the form which can be studied using the Holmes-Marsden modification of the Melnikov method. This simple modification extends the method from one to two degrees of freedom. There also exists a generalization to systems with still more degrees of freedom (see \citealt{Gruendler-85} or Chapter 4 of \citealt{Wiggins-88}), but that is much more complicated. In particular, it requires a complete solution of the variational equation of the unperturbed system along the homoclinic orbit, which is however not known analytically in most cases.

There also exist other generalizations of the original Melnikov method. One of them is to the case when neither the unperturbed system nor the perturbation are Hamiltonian in nature, which mainly involves dissipative systems (drag forces, friction, etc.); see e.g. \cite{Wiggins-03} for examples. Another variant is when the Melnikov integral is taken along a heteroclinic orbit (which approaches {\em different} hyperbolic-fixed-point orbits in the past and future).

Finally, we briefly mention the previous usage of Melnikov's method for test motion in perturbed black-hole fields. 
\cite{Moeckel-92} considered a geodesic flow in a central field under the influence of an additional distant body and showed it leads to the relativistic version of a classical Hill problem. \cite{BombelliC-92} considered a periodic perturbation (e.g. due to gravitational waves); in that case the perturbed motion remains planar, so the authors did not need to generalize the method to more degrees of freedom. Geodesic chaos induced by perturbation of Schwarzschild due to gravitational waves was also studied in a similar manner by \cite{LetelierV-97}; see also \cite{LetelierV-98} where they analysed, using the Melnikov method, the periodically perturbed equatorial motion around a centre modelled by an inverse-square law plus a quadrupole-like term.

\section{Concluding remarks}
\label{concluding}

Motivated by theoretical interest as well as by accreting black holes in astrophysics, we study the time-like geodesic dynamics in space-times of static black holes perturbed by a ring or a disc. In the present paper of the series, we have used the Melnikov method which detects whether, under perturbation, the homoclinic orbit of the originally fully integrable system (geodesic flow in the field of the black hole alone) breaks up into transversally intersecting stable and unstable asymptotic manifolds, which implies the occurrence of chaotic behaviour. We considered, specifically, the Schwarzschild black hole simulated by the Nowak-Wagoner pseudo-potential and encircled by the Bach-Weyl ring or the inverted first Morgan-Morgan disc, and the extreme Reissner-Nordstr\"om black hole encircled by the extremally charged, Majumdar-Papapetrou ring.

In order for the Melnikov method to be applicable to our systems (with two degrees of freedom), we made a canonical transformation of the respective Hamiltonians, and used the results of \cite{HolmesM-83}.
For all our systems the Melnikov function was found to have simple zeros which proves that the homoclinic orbit really breaks up into a chaotic layer. In agreement with the Melnikov theory, for a small perturbation the chaotic layer only covers a small part of some hypersurfaces in the phase space, given by values of the integrals of motion for which the original homoclinic orbit (separatrix) indeed exists. We verified the results obtained by the Melnikov method numerically for the electro-vacuum Majumdar-Papapetrou space-time generated by the extreme Reissner-Nordstr\"om black hole encircled by the extremally charged ring. It can be expected that for superpositions with the Schwarzschild black hole the results would be similar.

The usage of canonical transformation for putting the Hamiltonian into a form suitable for the Melnikov method is not restricted to our particular systems -- similar approach could actually be applied to a stationary central field with any axially symmetric perturbation. Such a technique is quite simple in comparison with the generalizations of the Melnikov method to more degrees of freedom considered in the literature.

\begin{acknowledgments}
We thank for support from the grants GACR-17/06962Y (L.P.) and GACR-17/13525S (O.S.) of the Czech Science Foundation.
And we are grateful to O. Kop\'a\v{c}ek, G. Lukes-Gerakopoulos and P. Sukov\'a for helpful discussions.
\end{acknowledgments}

\bibliography{chaotic6.bib}

\end{document}